\documentclass[11pt,twocolumn]{emulateapj}

\tightenlines
\usepackage{epstopdf}
\usepackage{soul} % DCW, alternative to ulem that allows line breaks
\usepackage{color} % AV
\usepackage{lineno}
\usepackage{amsmath}
\usepackage{lipsum} % for testing only

\def\lsim{\;\raise0.3ex\hbox{$<$\kern-0.75em\raise-1.1ex\hbox{$\sim$}}\;}
\def\gsim{\;\raise0.3ex\hbox{$>$\kern-0.75em\raise-1.1ex\hbox{$\sim$}}\;}

\definecolor{purple}{RGB}{200,100,255} %{255,100,20}

\newcommand{\gamZ}{\gamma_0}

\newcommand{\xx}[1]{\!\times\!10^{#1}}

\newcommand{\epse}{\epsilon_{e}}
\newcommand{\epsB}{\epsilon_{B}}

\renewcommand{\baselinestretch}{1}
\def\aap{Astronomy and Astrophysics}
\def\apj{Astrophysical Journal}
\def\apjl{The Astrophysical Journal Letters}
\def\mnras{MNRAS}

\def\nar{New Astronomy Reviews}

\newcount\listnorom
\newcommand\listromanDE{\global\advance \listnorom by 1
{\lowercase\expandafter{(\romannumeral\listnorom)}\ }}

\newcount\listnumber
\newcommand\listDE{\global\advance \listnumber by 1
{\lowercase\expandafter{(\number\listnumber)}\ }}

\newcount\Inum
\newcount\IInum
\newcount\IIInum
\newcount\IVnum
\def\I{\global\multiply\IInum by 0 \global\multiply\IIInum by 0
            \global\multiply\IVnum by 0 \global\advance \Inum by 1
            {\the\Inum. }}
\def\II{\global\multiply\IIInum by 0\global\multiply\IVnum by 0
       \global\advance \IInum by 1 {\the\Inum.\the\IInum. }}
\def\III{\global\multiply\IVnum by 0\global\advance \IIInum by 1
            {\the\Inum.\the\IInum.\the\IIInum. }}
\def\IV{\global\advance \IVnum by 1
            {\the\IVnum. }}
\shorttitle{SSA in afterglows with thermal particles}
\shortauthors{Warren et al.}

\begin{document}
%\medskip

\title{Synchrotron self-absorption in GRB afterglows: the effects of a thermal electron population} 

\vskip24pt

\author{Donald C. Warren,\altaffilmark{1}
Maxim V. Barkov,\altaffilmark{1,2}
Hirotaka Ito,\altaffilmark{1}
Shigehiro Nagataki,\altaffilmark{1,3-5}
Tanmoy Laskar,\altaffilmark{6,7}
}

\altaffiltext{1}{Astrophysical Big Bang Laboratory, RIKEN, Wako, Saitama 351-0198, Japan; donald.warren@riken.jp}
\altaffiltext{2}{Department of Physics and Astronomy, Purdue University, 525 Northwestern Avenue, West Lafayette, IN 47907-2036, USA}
\altaffiltext{3}{Interdisciplinary Theoretical Science (iTHES) Research Group, RIKEN, Wako, Saitama 351-0198, Japan}
\altaffiltext{4}{Interdisciplinary Theoretical \& Mathematical Science (iTHEMS) Program, RIKEN, Wako, Saitama 351-0198, Japan}
\altaffiltext{5}{College of Science, Department of Physics, Rikkyo University, Nishi-Ikebukuro, Tokyo 171-8501, Japan}
\altaffiltext{6}{National Radio Astronomy Observatory, 520 Edgemont Road, Charlottesville, VA 22903, USA}
\altaffiltext{7}{Department of Astronomy, University of California, 501 Campbell Hall, Berkeley, CA 94720-3411, USA}

%%%
\begin{abstract}
%%%

In the standard synchrotron afterglow model, a power law of electrons is responsible for all aspects of photon production and absorption.  Recent numerical work has shown that the vast majority of particles in the downstream medium are actually ``thermal'' particles, which were shock-heated but did not enter the Fermi acceleration process (the name stands in contrast to the nonthermal high-energy tail, rather than connoting a Maxwellian distribution).  There are substantial differences at optical and higher energies when these thermal electrons participate in the afterglow, but early work along these lines ignored the radio end of the electromagnetic spectrum.  We report here on an extension of previous Monte Carlo simulations of gamma-ray burst afterglows.  The model now includes the synchrotron self-absorption (SSA) process and so can simulate afterglows across the entire EM spectrum, and several orders of magnitude in time.  In keeping with earlier work, inclusion of the thermal electrons increases the SSA frequency by a factor of 30, and the radio intensity by a factor of 100.  Furthermore, these changes happen with no modification to the late optical or X-ray afterglow.  Our results provide very strong evidence that thermal electrons must be considered in any multiwavelength model for afterglows.

\end{abstract}

%%%
\section{Introduction}
\label{sec:intro}
%%%

Gamma-ray bursts (GRBs) are among the most energetic events since the birth of the Universe, and understanding them offers insight into physics far beyond anything testable in terrestrial laboratories.  The first detection of the afterglows of long GRBs occurred two decades ago, and revolutionized the field by settling the debate surrounding their cosmological origin; by yielding a firm association with stripped-envelope core-collapse supernovae; and by enabling a search for their progenitors to determine their energy scale, circumburst medium, and degree of collimation, along with a precise localization within their host galaxies \citep{vanParadijs_etal_1997, Costa_etal_1997, Frail_etal_1997, PanaitescuKumar2001ApJ554, WoosleyBloom2006, BBF2016}. Similarly, the detection of short GRB afterglows following the launch of \textit{Swift} has enabled detailed studies of their energetics and explosion environments, supporting the compact binary progenitor hypothesis \citep{Blinnikov_etal_1984, Paczynski1986, FBF2010, BFC2013}.  This connection was, of course, confirmed by the near-simultaneous detections of GW 170817 by the LIGO/VIRGO collaboration and of GRB 170817A by \textit{Fermi} and numerous others \citep{Abbott_etal_2017PhRvL119p, Goldstein_etal_2017, Abbott_etal_2017ApJ848L}; and see \citet{Pozanenko_etal_2018} for an extended discussion of how the physical properties of the event are constrained by multiwavelength observations.

Given the importance of radio observations to the study of GRB afterglows, properly understanding their production is critical.  The traditional model for afterglow emission assumes that all electrons are accelerated to a power-law distribution, $N_{e}(E) \propto E^{-p}$, with a low-energy cutoff $E_\mathrm{min}$ \citep{SPN1998}.  Far fewer works have explored the possibility that electrons \textit{don't} (or don't \textit{just}) form a power law.  \citet{EichlerWaxman2005} calculated the observational consequences of incomplete acceleration.  In their model only a fraction $f$ of the electrons are accelerated into a power law distribution, with the remainder forming a quasi-thermal distribution at an energy lower by a factor $\approx \eta m_{e}/m_{p}$ than $E_\mathrm{min}$, where $\eta$ is a free parameter; these low-energy electrons do not contribute significant emission.  Afterglow models are completely degenerate with respect to the participation parameter $f$, but see \citet{TIN2008} for a discussion of breaking the degeneracy using polarization.  Prompted by later numerical results \citep[such as][]{Spitkovsky2008}, \citet{GianniosSpitkovsky2009} studied the early X-ray afterglow when the electron population is split into a thermal distribution at the base of a power-law tail.  The changing curvature of such an electron distribution causes a non-monotonic hard-soft-hard variation in the photon spectral index as the characteristic synchrotron frequency of the thermal peak passes through a given energy band.  Eliminating the power law completely, and placing all the electrons in a thermal distribution, would produce a sharp drop in photon production between the synchrotron spectrum and its Comptonized echo at higher energies \citep{PVP2014}.

\citet{WEBN2017} simulated GRB afterglows by self-consistently modeling Fermi acceleration at a relativistic shock.  A significant aspect of that work was the emphasis that the electron population in the downstream, shocked, region is not just a nonthermal power-law distribution.  There is an additional ``thermal'' population of electrons that were shock-heated but not injected into the Fermi acceleration process.\footnote{``Thermal'' is used here as a contrast to the ``nonthermal'' power law distribution.  We make no claims about whether the shock-heated particles form a Maxwellian distribution.}  Including the downstream thermal particles in photon production processes makes a substantial difference in the temporal evolution of afterglow luminosity and spectra, as does adjusting the efficiency with which downstream particles enter the acceleration process.  That work considered only photons at optical or higher energies, deferring the radio afterglow to a later paper.

Most recently, \citet{ResslerLaskar2017} computed the observational effects of including a population of non-accelerated electrons in an idealized, spherical, relativistic GRB jet.  By integrating the radiative transfer equation over equal arrival time surfaces, they derived model spectra and light curves accounting for synchrotron cooling and self-absorption effects.  They found that the inclusion of non-accelerated electrons increases the optical depth of self-absorption in the centimeter bands, while generating detectable emission in excess of the synchrotron radiation from power-law electrons in the millimeter and optical frequencies.

Here, we expand on the results of \citet{WEBN2017}, extending that work to include photon production/absorption processes at all energies.  Specifically, we have implemented synchrotron self-absorption (SSA) by arbitrary electron distributions.  We provide only the barest description of our model here, in Section~\ref{sec:model}; for details see the extensive discussion in \citet{WEBN2017}.  In Section~\ref{sec:SSA} we explain the new addition to our simulated afterglows.  In Section~\ref{sec:results} we place these absorbed afterglows in the context of observations, and we conclude in Section~\ref{sec:conclusions}.

%%%
\section{Our model}
\label{sec:model}
%%%

Our simulated afterglows use the \citet{BlandfordMcKee1976} solution for both the motion of the shock front and all shocked fluid behind it.  We divide our shock evolution into several time steps.  During each time step, we simulate the acceleration of swept-up ions and electrons by a shock whose Lorentz factor is prescribed by the Blandford--McKee solution.  Swept-up plasma from previous time steps is arranged in shells behind the shock \citep[Figure~1 of][]{WEBN2017}, and we allow all shocked shells to participate in photon production at all time steps.  We consider three different photon production processes---synchrotron, inverse Compton, and pion decay from hadronic collisions---but only synchrotron is relevant to this work.

Both adiabatic (for all species) and radiative (for electrons only) cooling takes place during each time step.  We assume a frozen-in magnetic field, so the field strength also decays as the fluid moves downstream from the shock.  Note that we do not consider the decay of amplified microturbulence, which would cause much faster decay over much shorter length scales: see \citet{Lemoine2013} for a discussion of the observational consequences of a quickly-decaying magnetic field.  Since radiative cooling may be extremely rapid just behind the shock, the most recent time step is divided into sub-intervals for the purposes of cooling and photon production; all such sub-intervals follow their own, independent, cooling history.  These sub-intervals, their widths, their magnetic field strengths, and their electron spectra are also used to compute the effects of synchrotron self-absorption.

As in \citet{WEBN2017}, the hydrodynamic parameters we assume are an isotropic explosion energy $E_\mathrm{iso} = 10^{53}$~ergs, and a constant-density circumburst medium with $n_{0} = 1$~cm$^{-3}$.  For the purposes of calculating photon production, we assume that $\epse = 0.35$ and $\epsB = 10^{-3}$, where $\epsilon$ is the fraction of the total energy density in the local plasma frame, just behind the shock.  The subscripts $e$ and $B$ refer to electrons and the magnetic field, and the remainder of the energy density is given to ions (i.e. $\epse + \epsB + \epsilon_{i} \equiv 1$).  We further assume a comoving distance of 1~Gpc for the source, corresponding to a redshift $z = 0.23$.

%%%
\section{Synchrotron self-absorption}
\label{sec:SSA}
%%%

The traditional treatment of SSA comes from \citet{SPN1998} and \citet{GPS1999ApJ527}.  The ``back of the envelope'' calculation for the self-absorption frequency $\nu_{a}$, where the optical depth becomes unity, is \citep[for a constant-density CBM;][]{GPS1999ApJ527}
\begin{align}
  \nu_{a} &= 4.24\xx{9} (1+z)^{-1} \left(\frac{p+2}{3p+2}\right)^{3/5}  \nonumber \\
              &\quad \times \frac{(p-1)^{8/5}}{p-2} \epse^{-1} \epsB^{1/5} E_{52}^{1/5} n_{0}^{3/5}~\mathrm{Hz}
  \label{eq:GPS99_nua}
\end{align}
in the observer frame.  In the rest frame of the absorbing plasma, the redshift may be dropped and the Lorentz boost between that frame and the local ISM frame must be removed.  For a relativistic shock with a Lorentz factor $\gamZ$ and a shock-frame velocity ratio $u_\mathrm{UpS}/u_\mathrm{DwS}=3$, this boost is $\gamZ/\sqrt{2}$.  With these changes, and with the GRB parameters listed in Section~\ref{sec:model}, this equation becomes
\begin{equation}
  \nu_{a} = 3.41\xx{9} \gamma_{0}^{-1} \left(\frac{p+2}{3p+2}\right)^{3/5} \frac{(p-1)^{8/5}}{p-2}~\mathrm{Hz} .
  \label{eq:GPS99_nua_2}
\end{equation}
Note that, unlike in Equation~\ref{eq:GPS99_nua}, in the rest frame of the absorbing plasma $\nu_{a}$ increases with time.

SSA in our model is more complicated for several reasons.  First, we include the possibility that radiative losses occur and influence the SSA process.  Second, we cannot assume a particular shape for our electron distributions, either power-law or Maxwellian: the distribution of absorbing electrons arises self-consistently out of our Monte Carlo simulations, rather than being prescribed by an analytical formula. Finally, our model for the downstream plasma is multi-zone \citep[see Figure~1 of][]{WEBN2017}, which means multiple absorbing populations in different reference frames \citep[and which invalidates the assumption of isotropy made for the one-zone derivation in][]{RybickiLightman1979}.  We discuss each of these complications, and our treatment of them, below.

The phase space distribution of electron population downstream from the shock front is not stationary.  High-energy electrons cool as they radiate in the compressed and amplified magnetic field, piling up towards the low-energy end of the distribution.  While the shock is ultrarelativistic, the extreme magnetic fields may produce cooling even for the lowest-energy electrons.  Our simulations take place in the slow-cooling regime, but this merely means that the electrons do not immediately radiate away their entire energy.\footnote{SSA occurring well into the fast cooling regime was studied by \citet{GPS2000}, again for a pure power-law electron distribution.  The transition between fast and slow cooling occurs at \citep{SPN1998}
\begin{align*}
  t_\mathrm{trans} &= 210 \epsB^{2} \epse^{2} E_{52} n_{1}~\mathrm{days} \\
     &= 22~\mathrm{sec} ,
\end{align*}
for our fiducial parameters, well before the beginning of our simulations.}  We cannot, and do not, assume that the lowest-energy electrons are unaffected by radiative cooling at all times.

Since our electron distributions are neither purely thermal nor purely power-law, we cannot use either of the simple closed-form solutions presented in \citet{RybickiLightman1979}.  Instead, we must begin with their Equation~(6.50),
\begin{equation}
  \alpha_{\nu} = -\frac{c^{2}}{8 \pi \nu^{2}} \int P(\nu,E) \cdot E^{2} \frac{d}{dE} \left( \frac{1}{E^{2}} \frac{dn}{dE} \right) dE .
  \label{eq:absorption_coeff}
\end{equation}
which expresses the absorption coefficient $\alpha_{\nu}$ at frequency $\nu$ in terms of the electron energy distribution $dn/dE$ \citep[called $N(E)$ in][]{RybickiLightman1979}, and the synchrotron power emitted at that frequency by an electron of energy $E$.  When $\nu \ll \nu_{c}(E_\mathrm{min})$ for the lowest-energy electrons, $P(\nu,E)$ may be rewritten
\begin{equation}
  P(\nu) = \frac{2^{5/3} \pi q^{3} B \mathrm{sin}\alpha}{m_{e} c^{2} \Gamma(\frac{1}{3})} \left(\frac{\nu}{\nu_{c}}\right)^{1/3},
  \label{eq:Pnup_rewrite}
\end{equation}
where
\begin{equation}
  \nu_{c} = \frac{3 \gamma^{2} q B \mathrm{sin} \alpha}{4 \pi m_{e} c}
  \label{eq:nu_char}
\end{equation}
is the characteristic frequency of synchrotron emission by electrons with charge $q$ and Lorentz factor $\gamma$ in a local magnetic field of strength $B$, with a pitch angle $\alpha$ between the electron's momentum and the magnetic field orientation.

The optical depth through a source is defined \citep[][Equation 1.26]{RybickiLightman1979} as
\begin{equation}
  \tau_{\nu} = \int_{s_{0}}^{s} \alpha_{\nu}(s^{\prime}) ds^{\prime} ,
  \label{eq:opt_depth}
\end{equation}
where $s$ is the (comoving) position within each shell, and the absorption coefficient $\alpha_{\nu}$ is defined above in Equation~\ref{eq:absorption_coeff}.  Fortunately, it is a Lorentz invariant, which allows us to calculate the optical depth using the following process.  (1) First, in each emitting region we transform photon energies into the reference frame of each absorbing region ahead of it.  (2) Using the transformed photon energy and the local electron distribution---which we assume to be isotropic---we calculate the optical depth due to electrons in the absorbing region. (3) Finally, we sum all contributions to the optical depth from regions ahead of the emitting region.  As mentioned in Section~\ref{sec:model}, sub-intervals of the most recently shocked plasma (where radiative cooling is most important) are handled independently from each other, rather than assuming a single plasma state for the entire shell.

Because we consider absorption by multiple shells, which may have significant Lorentz boosts relative to each other, and which have different electron distributions (and so different values for $\alpha_{\nu}$), we cannot present a closed-form solution for the optical depth.  Instead we calculate $\tau_{\nu}$ for each frequency, over all shocked shells and sub-intervals, and identify the self-absorption frequency $\nu_{a}$ as that where $\tau_{\nu} = 1$.  (Examples of this analysis are presented in Figure~\ref{fig:nua_by_model} later.)

\subsection{Analytical estimate of SSA}

Assume (for the moment) that the electron distribution is both isotropic and a simple power law, i.e.,
\begin{equation}
  \frac{dn}{dE} = n_\mathrm{PL} (p-1) E_\mathrm{min}^{p-1} E^{-p} ,
  \label{eq:dndE_subs}
\end{equation}
where $n_\mathrm{PL}$ is the total density of power-law electrons as measured in their local frame.  After substituting Eqs.~\ref{eq:Pnup_rewrite},\ref{eq:nu_char} and \ref{eq:dndE_subs} into Equation~\ref{eq:absorption_coeff}, one finds that
\begin{equation}
  \alpha_{\nu,\mathrm{PL}} = g(p) \frac{3^{2/3} \pi^{5/6} q^{8/3} c^{5/3}}{2^{2/3} \, 5 \, \Gamma(\frac{5}{6})} B^{2/3} n_\mathrm{PL} E_\mathrm{min}^{-5/3} \nu^{-5/3} ,
  \label{eq:absorption_coeff_PL}
\end{equation}
where $g(p) \equiv (p-1)(p+2)/(3p+2)$ contains the dependence on the spectral index of the power law, and where the pitch angle has been averaged out: $\langle \mathrm{sin}^{2/3} \alpha \rangle = \sqrt{\pi} \, \Gamma(1/3) / (5 \, \Gamma(5/6))$.

The electrons that were shock-heated but not Fermi-accelerated will also contribute to SSA.  As demonstrated in Figure~2 of \citet{WEBN2017}, this population may be treated as a delta function in energy, positioned at the base of the nonthermal tail, i.e.,
\begin{equation}
  \frac{dn}{dE} = n_\mathrm{DF} \delta(E - E_\mathrm{min}) ,
  \label{eq:dndE_subs2}
\end{equation}
where the subscript ``DF'' refers to the delta function shape assumed for the electron distribution.  In this case, repeating the calculations of Equations~\ref{eq:absorption_coeff}-\ref{eq:absorption_coeff_PL}, and recalling the identity $f(x)\delta^{\prime}(x-a) = -f^{\prime}(x)\delta(x-a)$, yields
\begin{equation}
  \alpha_{\nu,\mathrm{DF}} = \frac{2^{4/3} \pi^{5/6} q^{8/3} c^{5/3}}{3^{4/3} \, 5 \, \Gamma(\frac{5}{6})} B^{2/3} n_\mathrm{DF} E_\mathrm{min}^{-5/3} \nu^{-5/3} ,
  \label{eq:absorption_coeff_DF}
\end{equation}
which bears a strong resemblance to Equation~\ref{eq:absorption_coeff_PL}. Indeed, the ratio of the two absorption coefficients is
\begin{equation}
  \frac{\alpha_{\nu,\mathrm{PL}}}{\alpha_{\nu,\mathrm{DF}}} = \frac{9}{4} \frac{ (p+2)(p-1) }{ 3p+2 } \frac{ n_\mathrm{PL} }{ n_\mathrm{DF} } .
  \label{eq:absorption_coeff_ratio}
\end{equation}
The numerical prefactor is of order unity for $2 \le p \le 5$, and so the density ratio of the two populations is a much larger factor in their relative absorption.  PIC simulations show that the power-law part of the electron distribution contains, approximately, just a few percent of the total downstream electron population, meaning $n_\mathrm{DF} \gtrsim 30 \, n_\mathrm{PL}$.  The absorption due to the two groups of electrons is thus influenced far more strongly by the thermal population, a conclusion borne out in the next section.

%%%
\section{Results}
\label{sec:results}
%%%

In this section we apply Equations~\ref{eq:absorption_coeff} \& \ref{eq:opt_depth} to the three models presented in \citet{WEBN2017} to discuss the relevance of SSA to each.  These models are (1) a model where only the nonthermal particles accelerated at a test-particle shock are allowed to participate in photon production \citep[called here NT-only, and called CR-only in][]{WEBN2017}, in keeping with the standard synchrotron afterglow scenario; (2) a test-particle case (TP) where the shock is assumed to be a discontinuity in velocity; and (3) a fully nonlinear model (NL), where backpressure due to accelerated particles is allowed to modify the shock profile and the associated particle distributions.  See Figure~2 of \citet{WEBN2017} for the electron distributions as functions of both energy and time, as well as Figure~4 for the photon spectra without SSA.  To approximately conserve fluxes across the shock, both the NT-only and TP models have had their injection rates reduced relative to the NL model.\footnote{As discussed in, e.g., \citet{EWB2013}, our thermal leakage injection model is too efficient across a discontinuous velocity profile: too many particles are injected into the Fermi acceleration process, and the momentum/energy fluxes downstream from the shock do not match those far upstream.  One method to deal with this violation of the Rankine--Hugoniot conditions is to smooth the velocity profile, introducing a precursor and producing the NL particle distributions mentioned above.  If a discontinuous shock must be assumed, however, the injection efficiency must be reduced to diminish flux imbalances associated with the thermal leakage scheme.}  The three models are distinguished solely by their treatments of microphysics at the shock front, and do not cause any changes to the hydrodynamics, which follow the Blandford--McKee solution in all cases.

\subsection{The effect of electron distribution on SSA}
\begin{figure}
  \epsscale{1.15}
  \plotone{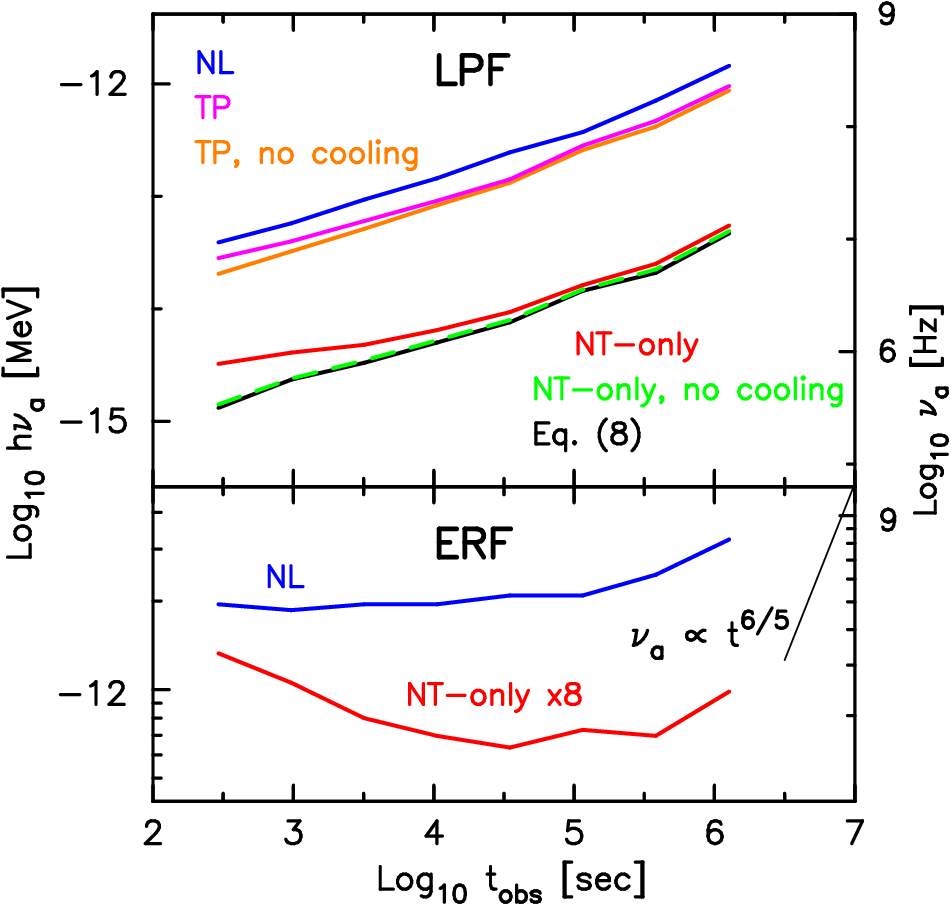}
  \caption{Synchrotron self-absorption frequency in the local plasma frame of the absorbing material (top frame), and Lorentz-boosted to the engine rest frame (for two of the models; bottom frame).  The curves represent the following scenarios: (black) $\nu_{a}$ as derived from Equation~\ref{eq:absorption_coeff_PL}; (green, dashed) NT-only spectrum, with no downstream evolution of the plasma (i.e., exactly the situation used to reach Equation~\ref{eq:absorption_coeff_PL}); (red) NT-only spectrum, but with cooling of energetic particles after they decouple from the shock; (orange) test-particle (TP) spectrum, with no cooling; (magenta) TP spectrum, with cooling; (blue) nonlinear (NL) spectrum, with cooling.  In the bottom frame, the thin guide line shows a $\nu \propto t^{6/5}$ dependence, appropriate for the non-relativistic Sedov--Taylor solution.}
  \label{fig:nua_by_model}
\end{figure}

We present the SSA frequency ($\nu_{a}$) as a function of observer time for several different scenarios in Figure~\ref{fig:nua_by_model}.  As mentioned in Section~\ref{sec:SSA}, the optical depth is computed for the entire shocked volume, not just the most recently shocked material.  In the top panel of Figure~\ref{fig:nua_by_model} we present $\nu_{a}$ in the local plasma frame of the material just downstream from the shock, before any Lorentz boost (into the engine rest frame) or redshift effects are applied---corresponding more closely to the result of Equation~\ref{eq:GPS99_nua_2} than that of Equation~\ref{eq:GPS99_nua}.  Observer time includes both of these effects, but is used as the abscissa for more convenient comparison with figures here and in \citet{WEBN2017}.  The curves in the top panel are not constant in time, as traditionally expected for observations of $\nu_{a}$, but this is because they are not shown in the observer frame.

The black curve in Figure~\ref{fig:nua_by_model} shows $\nu_{a}$ as calculated by Equation~\ref{eq:absorption_coeff_PL}, which is the simplest possible case for the multi-zone model we use: the photon calculations assume a constant state (pure power-law electron spectrum, local electron density, magnetic field strength) everywhere in each emission shell.  That is, the emission and absorption calculations ignore the downstream evolution as freshly-shocked electrons advect away from the shock.  (Cooling occurs between time steps to determine the correct state of each downstream emission shell, but these intermediate states are ignored for photon production and SSA.)  The green dashed curve in Figure~\ref{fig:nua_by_model} applies Equation~\ref{eq:absorption_coeff} to the particle spectra from \citet{WEBN2017}, under the same assumptions.  The two overlap almost exactly, from which we conclude that the method outlined in Section~\ref{sec:SSA} produces correct results when applied to the electron distributions generated by the Monte Carlo model.

The red curve shows the impact on the SSA process of magnetic field decay and radiative losses.  As expected, at earlier times (and so stronger magnetic fields in the downstream plasma for fixed $\epsB$), the contribution to optical depth of particles further downstream is more significant.  The radiative losses shift the electron distribution towards lower energy, while the decayed magnetic field reduces the characteristic energy of the synchrotron photons produced by the electrons; both of these effects increase the absorption.  At later times the energy density of the magnetic fields advected downstream has been reduced due to adiabatic expansion, and the freshly amplified magnetic field near the shock is weaker since the shock is slower.  The generally weaker magnetic fields reduce the importance of radiative losses and move the red curve closer to the black and green curves, which do not include cooling/decay.  For the times considered in this work, the consequences of magnetic field decay and radiative losses are mostly unimportant: even at $t_\mathrm{obs} = 300$~s, the difference between the red and black curves is a factor of 2.3, and that factor shrinks rapidly with time.  Given that most long-wavelength observations of GRB afterglows take place much later than 300~s, the impact of these processes on SSA may be safely ignored.

Ours is not the first paper to consider the downstream evolution of synchrotron-relevant quantities.  \citet{TolstovBlinnikov2003} computed light curves and spectra based on a power-law electron distribution that evolved according to the Blandford--McKee solution (similar to the ``NT-only'' curve in Figure~\ref{fig:nua_by_model} above).  They also predicted a self-absorption frequency that varied with time in the observer frame, a topic to which we return below.  In their work, as opposed to ours, this downstream evolution causes estimates of $\nu_{a}$ to be off by a factor of a few. For the GRB parameters used in that paper, $\nu_{a}$ should have been 4~GHz in the observer frame; instead it was approximately 10~GHz until an observer time of $10^{6}$~s, at which time it began to decrease.

While synchrotron cooling may be ignored in treating SSA, the contribution of the thermal particles may not.  The upper three curves in Figure~\ref{fig:nua_by_model} show the value $\nu_{a}$ as computed for the two models with thermal particles included.  Respectively, orange, magenta, and blue correspond to the TP model without cooling, the TP model with cooling, and the NL model with cooling (no curve is shown for the NL model without cooling as it would not be qualitatively different from the TP model without cooling).  It is immediately clear that the thermal particles have a large effect on the value of $\nu_{a}$, by virtue of their sheer number compared to the more energetic particles in the nonthermal tail---see Equation~\ref{eq:absorption_coeff_ratio} and discussion surrounding it.  Including thermal particles increases $\nu_{a}$ by a factor of $\approx 25$, similar to (but larger than) the change found by \citet{ResslerLaskar2017}.  The NL model shows slightly higher absorption frequencies than the TP model for two reasons.  First, it is a well-known effect of nonlinear shock modification that the downstream thermal peak occurs at a slightly lower energy than in unmodified shocks, even at relativistic speeds \citep{EWB2013}.  Second, nonlinear shocks produce a bridge of particles connecting the thermal peak to the nonthermal tail \citep[see Figure~2 in][]{WEBN2017}.  Both of these effects combine to place more electrons at lower energies in the NL model than in the TP model.

We note here that blindly applying Equation~\ref{eq:GPS99_nua} to the upper curves (TP and NL models) in Figure~\ref{fig:nua_by_model} yields incorrect results.  If solving for $\epse$ using a known $\epsB$ and $\nu_{a}$, one obtains values greater than 1.  Or, if taking the values known from the Monte Carlo simulation ($\epse = 0.35$ and $\epsB = 10^{-3}$), the computed value of $\nu_{a}$ will be 30-75 times higher than that calculated from Equation~\ref{eq:absorption_coeff}.  These errors are even larger than the factors of $\approx 5$ in $\epse$ observed by \citet{ResslerLaskar2017} when they fixed their participation factor $f_\mathrm{NT} = 0.2$.  Our nonthermal particles are even fewer than 20\% of the total downstream electron population, and we see a larger discrepancy between the ``true'' values and the calculated values of $\nu_{a}$ and $\epse$.

The bottom panel of Figure~\ref{fig:nua_by_model} shows $\nu_{a}$ after a Lorentz boost to the rest frame of the central engine (but still without redshift effects).  We also restrict the number of models to just two (NT-only and NL) for the purpose of clarity.  For a self-consistent simulation of particle acceleration and photon production, $\nu_{a}$ is not perfectly constant in the observer frame. Both the NT-only and NL models show an increase in $\nu_{a}$ at late times, as the shock transitions from the relativistic Blandford--McKee solution to the non-relativistic Sedov--Taylor solution. In the non-relativistic limit, the self-absorption frequency does vary with time: $\nu_{a} \propto t^{6/5}$ \citep[][Eq. 60]{Gao_etal_2013}.  The shock is still far from non-relativistic at the final time considered---indeed, $\gamZ \approx 2.7$---and comparison against the provided guide line shows that the $\nu_{a}(t)$ curves are still far from the Sedov--Taylor limit.  However, it is clear that just a few days into the afterglow (for our fiducial parameters), the self-absorption frequency is already departing from the relativistic solution.

\subsection{Comparison of self-absorbed spectra}
\label{sub:compare}
\begin{figure*}
  \epsscale{1.15}
  \plotone{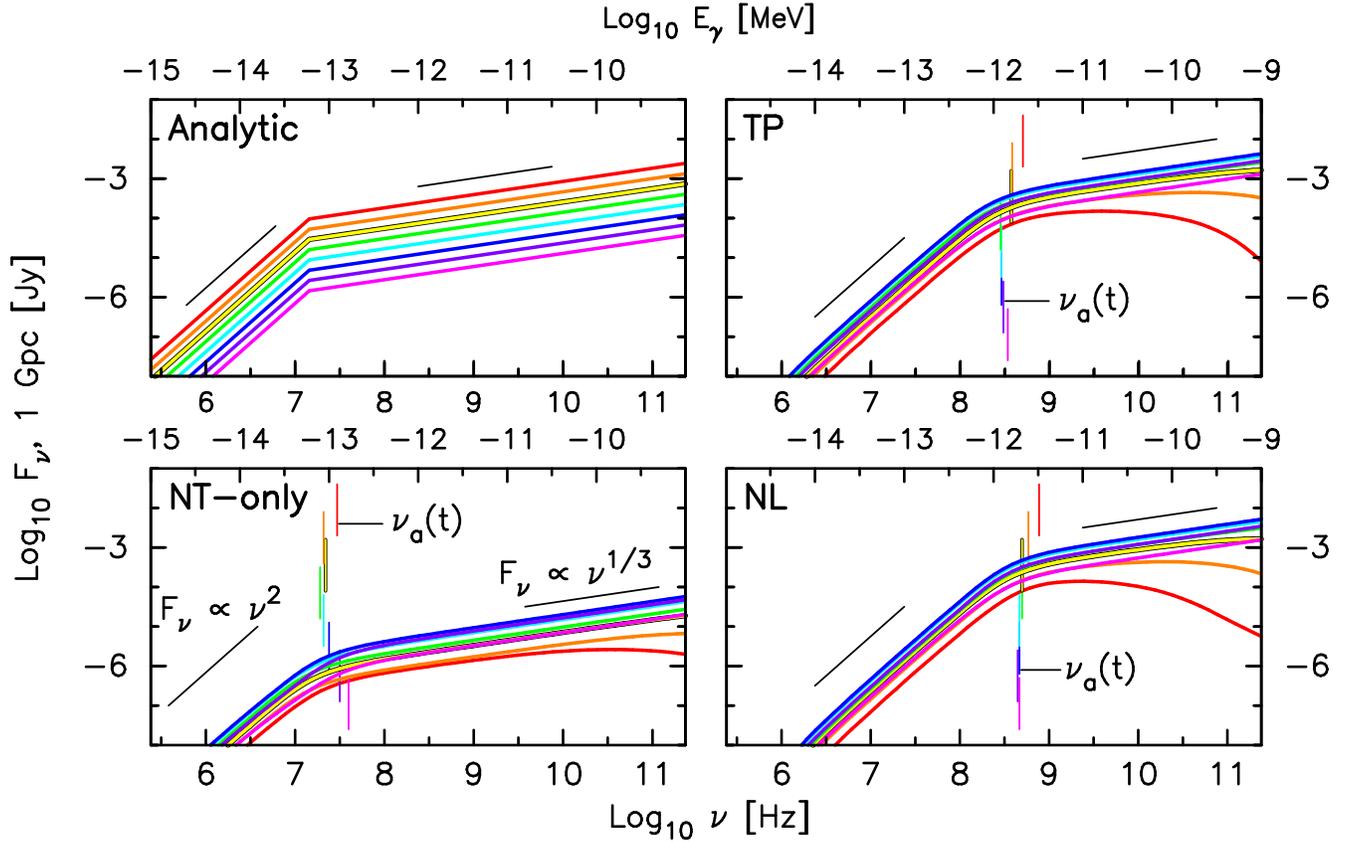}
  \caption{Low-frequency portion of photon spectrum, with SSA included, for four models.  In all four panels, thick curves trace the flux density at various observer times; the colors are described at the start of Section~\ref{sub:compare}.  Two thin guide lines in each panel show an $F_{\nu} \propto \nu^{2}$ and $F_{\nu} \propto \nu^{1/3}$ dependence.  In all but the first panel, vertical lines illustrate the location of the synchrotron self-absorption frequency $\nu_{a}$ as a function of time.  \textbf{Top left:} analytical model for flux density, described in Appendix~\ref{app:analytical}, with participation fraction $f = 0.01$. \textbf{Bottom left:} NT-only model, which includes only electrons that entered the Fermi acceleration process. \textbf{Top right:} TP model, in which accelerated particles have no effect on the shock structure, but in which thermal particles participate in all photon processes.  \textbf{Bottom right:} NL model, which adds the nonlinear interaction between Fermi-accelerated particles and the shock structure.}
  \label{fig:lowE_spectra_anCRTPNL}
\end{figure*}

The effect of SSA on photon spectra is illustrated in Figure~\ref{fig:lowE_spectra_anCRTPNL}.  The top left panel is a set of analytical spectra, explained in Appendix~\ref{app:analytical}.  GRB parameters are the same as for the other three models ($E_\mathrm{iso} = 10^{53}$~erg, $n_{0} = 1$~cm$^{-3}$, $\epse = 0.35$, $\epsB = 10^{-3}$), but a new participation fraction $f$ is included, with a value of $10^{-2}$.  The times span $t_\mathrm{obs} = 300$~s (magenta) to $t_\mathrm{obs} = 15$~d (red); the intermediate colors fall between those two extremes, with roughly a factor of three separating the values of $t_\mathrm{obs}$ at successive time steps.

All but one panel in Figure~\ref{fig:lowE_spectra_anCRTPNL} contains a sequence of vertical lines showing the time-dependence of $\nu_{a}$ for the particular model (the location of $\nu_{a}$ is much more obvious in the analytical case). The vertical placement of these lines is for convenience only, to allow easier comparison.  As expected from the discussion in the previous subsection, both the analytical and NT-only models have significantly lower values of $\nu_{a}$ throughout the simulated afterglow.  At 10~MHz, $\nu_{a}$ is right at the edge of the radio window, and determinations of the break frequency are all but impossible from the ground.  Even for the LOFAR instrument \citep[with a nominal frequency range of 30-240~MHz;][]{vanHaarlem_etal_2013} such an afterglow would require stronger magnetic fields, larger upstream densities, or much more energetic GRBs---or a combination of all three---for the self-absorption break to be observable.  The TP and NL models, on the other hand, produce absorption frequencies of several hundred MHz without resorting to extreme parameter choices.

Contrary to the predictions of the standard afterglow model (Equation~\ref{eq:GPS99_nua}), the observed self-absorption frequency is not constant in time.  Indeed, it isn't necessarily monotonic, as illustrated in the bottom panel of Figure~\ref{fig:nua_by_model}.  All numerical models show that $\nu_{a}$ drops in time (albeit only slightly for the models with a thermal population), then rises.  The initial drop is due to cooling downstream from the shock.  As seen in Figure~\ref{fig:nua_by_model}, cooling causes an increase in $\nu_{a}$ relative to models that don't include it (which includes, of course, the standard synchrotron model for afterglows).  The increase is smaller for models with a thermal population because the high-energy electrons (those that cool the most) make up a smaller proportion of the total absorbing population.  As the shock slows and the downstream magnetic field falls, cooling becomes less important and the curve $\nu_{a}(t)$ flattens out as seen in the bottom panel of Figure~\ref{fig:nua_by_model}.

The rise in $\nu_{a}$ at late times, for all numerical models, is a consequence of the decreasing shock speed.  Once the shock ceases to be fully relativistic, several of the assumptions fail that were made in deriving Equations~\ref{eq:GPS99_nua}, \ref{eq:absorption_coeff_PL} and \ref{eq:absorption_coeff_DF}.  The most important of these is the assumption that $E_\mathrm{min} \propto \gamZ$: in the trans-relativistic regime the rest-mass energy of downstream particles is no longer negligible compared to their kinetic energy after being shock-heated.

As for the spectra themselves, most spectra presented in Figure~\ref{fig:lowE_spectra_anCRTPNL} show the expected $F_{\nu} \propto \nu^{1/3}$ behavior of synchrotron spectra below the thermal (or minimum-energy) peak.  Only the red curves of the NL and TP models (at $t_\mathrm{obs} = 15$~d) do not show this behavior.  The location of the thermal peak is so close to the location of the absorption break that several parts of the synchrotron spectrum overlap.  \citep[See the discussion of break sharpness in][]{GranotSari2002}.

\subsection{Comparison against observations}

We now turn from describing the results of our simulations to comparing them against observations of GRB afterglows in the radio band, choosing 8.5~GHz as our representative frequency.  In \citet{ChandraFrail2012}, this frequency was used in virtually all discussion of the radio behavior of afterglows, and we follow their example.

\begin{figure}
  \epsscale{1.15}
  \plotone{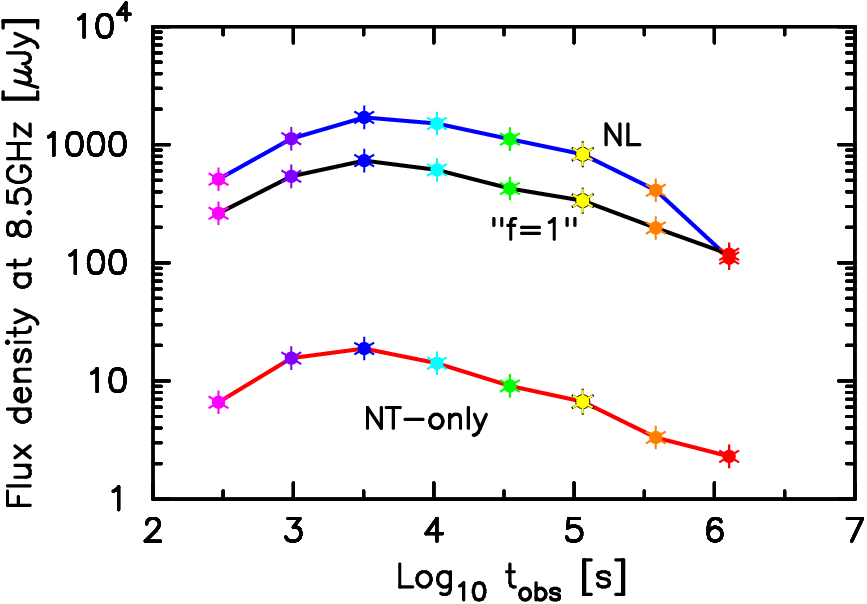}
  \caption{Light curve at 8.5~GHz for the NT-only (in red), NL (in blue), and ``$f=1$'' (black) models.  See text for a description of the last model.}
  \label{fig:8.5GHz_light_curve}
\end{figure}

\begin{figure}
  \epsscale{1.15}
  \plotone{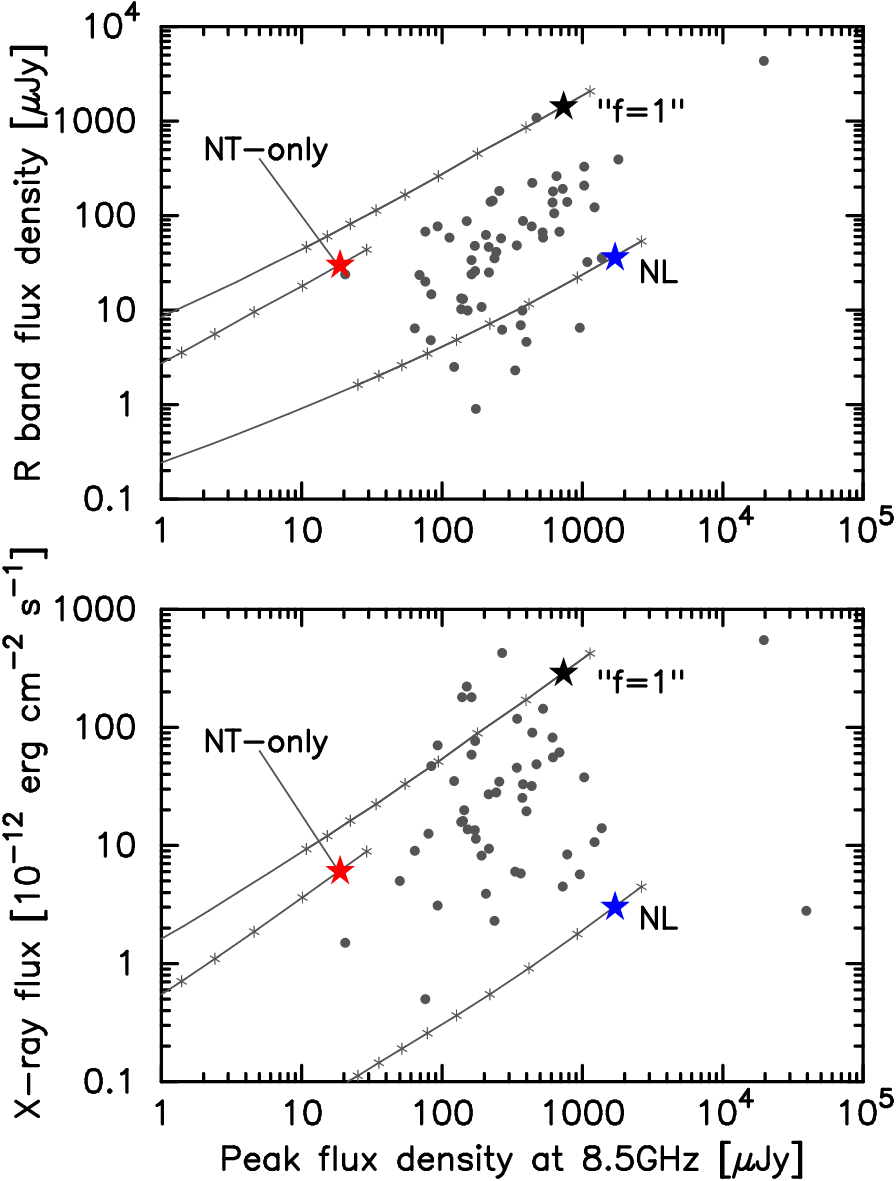}
  \caption{Scatterplot of X-ray and optical fluxes, at 11hr, plotted against peak 8.5~GHz flux density.  Solid dots are afterglow data taken from Tables~1 and 4 of \citet{ChandraFrail2012}, and the red, blue, and black stars show the NT-only, NL, and ``$f=1$'' models, respectively, at an assumed comoving distance of 1~Gpc.  The gray lines trace the redshift dependence of the three models; the small stars mark intervals $\Delta z = 0.1$, from $z = 0.2$ to $z = 1$.}
  \label{fig:CF2012_data}
\end{figure}

We present the 8.5~GHz light curves of the NT-only and NL models in Figure~\ref{fig:8.5GHz_light_curve}.  As expected from previous discussion, the NL model, which includes the thermal population of electrons, is dramatically brighter at radio wavelengths.  For the first seven time steps, the difference is roughly two orders of magnitude.  Only in the final time step, when the NL model shows a steepening due to the approach/passage of the thermal peak (see Figure~\ref{fig:lowE_spectra_anCRTPNL}), does the relative intensity begin to drop.  Of particular note is that this large increase in radio production comes with negligible change to optical and X-ray production at late times; we will return to this point later, using Figure~\ref{fig:CF2012_data}.

To underscore that the NT-only and NL models differ by more than a simple rescaling of the nonthermal tail, we also illustrate the light curve for a model we call ``$f=1$''. In this model we assume that all electrons fall within a nonthermal power-law distribution (i.e. the participation fraction $f = 1$). To achieve this we renormalize the electron distribution function of the NT-only model so that it has the same particle count as the proton distribution.  Such a distribution violates energy conservation across the shock, and is unphysical.  We \emph{do not} present it as a candidate for serious consideration, and emphasize this fact with the quotation marks around the model name. The NT-only and ``$f=1$'' light curves are identical, modulo their obvious rescaling and small deviations associated with the precise fraction of electrons in the thermal population at each time step.

The rise and fall of all three light curves, we stress, is not due to the passage of $\nu_{m}$, the characteristic synchrotron frequency of electrons in the thermal peak (or at the base of the nonthermal distribution).  It is clear from Figure~\ref{fig:lowE_spectra_anCRTPNL} that $\nu_{m}$ lies well above 8.5~GHz, for all models, for all but the final time step.  The non-monotonic behavior of the light curves is due instead to the changing conditions behind the shock as the Lorentz factor drops.  This behavior was also reported in \citet{TolstovBlinnikov2003}, whose 10~GHz light curve peaked at $t_\mathrm{obs} \approx 10^{4}-10^{5}$~s, at which point $\nu_{m}$ was greater than $10^{13}-10^{14}$~Hz.

The passage of $\nu_{m}$ does, however, have an observable effect on the light curves shown in Figure~\ref{fig:8.5GHz_light_curve}.  The final time steps of the NL model in Figure~\ref{fig:8.5GHz_light_curve} show a departure from the power-law decay after the light curve peaks.  This accelerated fading is due to the passage of $\nu_{m}$ across the 8.5~GHz band.  In Figure~\ref{fig:lowE_spectra_anCRTPNL}, both the TP and NL models start to deviate from the expected $F_{\nu} \propto \nu^{1/3}$ behavior above $\nu_{a}$ at late times, as $\nu_{m} \rightarrow \nu_{a}$.  This additional decrease is most noticeable for the red curves, at $t_\mathrm{obs} \approx 15$~d, and is stronger for the NL model because its thermal peak is at slightly lower energy than that of the TP model.  (The NT-only model does show this behavior as well.  The minimum energy of the NT-only model's electron spectrum is somewhat higher than those of the TP and NL models, though, and so $\nu_{m}$ is larger for this model at any given time.)

We note also that the difference between the ``$f=1$'' and NL light curves in Figure~\ref{fig:8.5GHz_light_curve} is a robust result, if not as extreme as the difference between the NL and NT-only models.  The NL model uses an electron distribution with a prominent thermal peak and a steep decay to the power-law nonthermal tail.  Despite being forced to contain the same number of electrons as the NL model, the ``$f=1$'' model is a pure power law.  Since the NL model's electron distribution is skewed towards lower energies, it naturally produces more emission at low photon energies than the ``$f=1$'' model does.  That said, once $\nu_{m}$ passes across any particular observed frequency, as is occurring at $t_\mathrm{obs} \approx 15$~d for the NL model, the situation reverses: since the ``$f=1$'' model does not have a thermal peak, its nonthermal tail contains more electrons than does the tail of the NL model.  After the passage of $\nu_{m}$, the ``$f=1$''  model will be brighter.

In Figure~7 of \citet{ChandraFrail2012}, light curves were given in both the observer frame and the engine frame for a set of well-observed cosmological GRBs.  The data appear to show a plateau in brightness to about $t_\mathrm{obs} = 10^{6}$~s (and possibly a two-peaked structure associated with reverse and forward shocks), and the authors attribute the subsequent drop in emission as the passage of $\nu_{m}$ for the forward shock of the GRB jet.

We posit an alternate explanation to that given in \citet{ChandraFrail2012}. Our self-consistent approach to particle acceleration and photon production produces 8.5 GHz light curves that peak well in advance of $10^{6}$~s (at which time $\nu_{m}$ is still above 8.5~GHz).  We also note that \citet{ChandraFrail2012} assumed that the microphysics parameters $\epse$ and $\epsB$ have not changed until $t_\mathrm{obs} \gtrsim 10^{6}$~s.  By this time, all but the most extreme afterglows\footnote{For a Blandford-McKee shock expanding into a homogeneous CBM, $\gamZ \gsim 10$ at $t_\mathrm{obs} = 10^{6}$~s implies $E_{52} \gsim 8\xx{4} n_{0}$.} should be well out of their relativistic phase of expansion.  Numerical simulations of relativistic shocks \citep{SSA2013,Ardaneh_etal_2015} show that the values of $\epse$ and $\epsB$ are well constrained by their origin in streaming instabilities ahead of the shock.  These instabilities saturate only in the relativistic regime, however, and may quench as the shocks slow below $\gamZ \approx 10$ \citep{LemoinePelletier2011}.  Without the instability to drive energy transfer from ions to electrons, the values of $\epse$ and $\epsB$ should change, invalidating one of the assumptions made in the standard synchrotron model for GRB afterglows.  Additional study is necessary to determine how $\epse$ and $\epsB$ change as the shock transitions from fully relativistic to nonrelativistic, and is beyond the scope of this work.

Multiwavelength comparisons are presented in Figure~\ref{fig:CF2012_data}, which uses data from Tables~1 and 4 of \citet{ChandraFrail2012}.  The optical and X-ray production of the NT-only and NL models are virtually identical at $t_\mathrm{obs} = 11$~h.  At this observer time, the high-energy emission comes from the nonthermal tail of the electron distribution, which is almost identical across these two models \citep[c.f. Figure~2 of][]{WEBN2017}.  As should be expected, the ``$f=1$'' model is significantly brighter at optical and X-ray wavelengths, since the nonthermal tail has been artificially enhanced relative to the two physically-motivated models.

Both the NT-only and NL models are too faint in X-rays at 11~h to easily match observed afterglows, but this is to be expected given the limitations of our current simulation setup.  Analytical and numerical work alike predicts significant, but localized, enhancement of the magnetic field in the immediate vicinity of the shock.  This will dramatically boost the production of X-ray photons, which must come from high-energy electrons that are either (1) still coupled to the shock, or (2) only recently decoupled and still mostly uncooled.  Both groups of electrons will be found near the shock, where the magnetic fields are strongest.  Optical and radio photons, on the other hand, come from lower-energy electrons throughout the shocked plasma, where the magnetic field is less affected by turbulence-driven amplification.  (While the NT-only and NL models are too faint in X-rays, the ``$f=1$'' model is brighter than the average observed afterglow.  However, we stress again that (1) this model is unphysical, and (2) despite a slightly better match to observed X-ray fluxes at 11~hr, we are not seriously proposing it to explain the observations.)

In contrast to shorter wavelengths, the peak flux density at 8.5~GHz, also seen in Figure~\ref{fig:8.5GHz_light_curve} above, depends sensitively on the number of low-energy particles.  We consider this difference in radio production to be extremely strong evidence for a thermal population below the nonthermal tail.  One cannot simply scale up the number of particles in the nonthermal distribution, as that will cause an increase in X-ray and optical production.  This is evident from comparing the NT-only and ``$f=1$'' redshift curves.  The spectral index of the electron distribution is constrained by the photon spectra, so one is not free to soften it arbitrarily and allow relatively more electrons at low energies compared to high.  Furthermore, numerical simulations show that the electron distribution has a fairly firm minimum energy around $\epse \gamZ m_{p} c^{2}$ due to energy transfer, which precludes merely extending the minimum energy of the nonthermal tail to the desired value.

Comparing the NL and ``$f=1$'' curves in Figure~\ref{fig:CF2012_data} illustrates the constraints given above.  While the ``$f=1$'' model is orders of magnitude brighter at X-ray and optical wavelengths (due to the greatly enhanced nonthermal electron population), it is fainter than the NL model at radio wavelengths because there are fewer electrons at the low-energy end of the distribution.  To make the ``$f=1$'' model a better fit to the observed data, one would want to increase the number of low-energy electrons (moving the curve right in both panels) while reducing the normalization of the nonthermal tail (moving the curve down).  However, the spectral index of the nonthermal tail is limited by observations at X-ray and optical frequencies, so choosing a different power law may not be possible. One would need a separate electron population at low energies that is independent of the nonthermal X-ray- and optical-producing electrons. In other words, the changes necessary to make the ``$f=1$'' model a better fit to the observations result in something qualitatively close to the NL model, with its populations of thermal and nonthermal electrons.

Given the wide range of parameters available to afterglow fits (including the redshift $z$, ambient medium profile $s$, ambient density $A_{\star}$/$n_{0}$, explosion energy $E_\mathrm{iso}$, peak Lorentz factor $\Gamma_\mathrm{pk}$, jet opening angle $\theta_{j}$, electron/magnetic field equipartition fractions $\epse$/$\epsB$, and electron spectral index $p$), it is certainly possible to fit each grey point in Figure~\ref{fig:CF2012_data} with either the NT-only model or the NL model.  As should be clear from Figure~\ref{fig:CF2012_data}, and from the results of \citet{ResslerLaskar2017}, the best-fit parameters will be dramatically different if one allows for a hot thermal population of electrons.  The difference between the inferred and true values can be as large as a factor of $f$ (or $1/f$). Given the small participation fractions expected from both PIC simulations and from our own Monte Carlo models \citep[$f \approx 0.01-0.03$, as seen in, e.g.,][and others]{SSA2013,WEBN2017}, afterglow models that include thermal electrons may result in a vastly different picture of GRB engines, progenitors, and environments.

We do note, also, that the standard model occasionally leads to ``best fits'' that are unlikely.  In both \citet{FWK2000} and \citet{Laskar_etal_2016}, the best fit parameters involved $\epse + \epsB > 0.95$, leaving just 5\% (or less) of the total energy density available to ions.  This is at odds with our present understanding of shock physics, and suggests that one (or more) of the assumptions made in the modeling must be revisited.

%%%
\section{Conclusions}
\label{sec:conclusions}
%%%

In this work we have presented radio emission from multiple models for the afterglows of gamma-ray bursts.  We focused on two models in particular.  Our NT-only model uses only a power-law distribution of nonthermal, shock-accelerated, electrons.  Our NL model adds a shock-heated (but \textit{not} further accelerated by the Fermi process) population of electrons, and additionally considers the back-reaction of accelerated particles on the shock doing the accelerating.

The presence of these thermal particles in the downstream plasma significantly increases the synchrotron self-absorption of the afterglow (Equation~\ref{eq:absorption_coeff_ratio} and Figure~\ref{fig:nua_by_model}).  This is because the absorption coefficients of the two populations (thermal and nonthermal) depend on their density, and numerical work shows that the nonthermal electrons are just a few percent of the total downstream population.  For reasonable choices of density, explosion energy, and magnetic field strength, the self-absorption frequencies $\nu_{a}$ of both the NT-only model and an analytic approximation (Figure~\ref{fig:lowE_spectra_anCRTPNL}) are too low to be detectable.  The contribution from thermal particles in the NL model pushes $\nu_{a}$ into observable frequencies.

Since the thermal population includes the lowest-energy electrons in the afterglow, it is unsurprising that radio emission would be boosted in the NL model compared against the NT-only model (Figure~\ref{fig:8.5GHz_light_curve}).  Slightly more surprising is that observations of radio afterglows show a plateau in emission, with a possible peak, far after our models predict that radio emission should be declining.  Our model cannot produce a radio peak at $t_\mathrm{obs} = 10^{6}$~s for any reasonable set of parameters.  The peak may be produced by a \textit{non}-relativistic shock, or by other physical effects like a counter-jet, but additional parameters must be introduced that cloud any claims of origin at present.

The boost that thermal electrons provide to radio emission comes with no change to their optical and X-ray production at late times; the nonthermal tail is responsible for the higher-energy photons then.  This is most apparent in Figure~\ref{fig:CF2012_data}, which compares our NT-only and NL models against observed afterglows.  In X-ray both models are slightly faint, but X-ray production will rely heavily on the microphysics near the shock, as both the highest-energy electrons and the most intense magnetic fields should be located there \citep[e.g.,][]{Lemoine2013}.  The interaction between microturbulence and thermal particles, and the consequences to observed photon spectra and light curves, will be explored in future work.

We used the following fiducial parameters for the large-scale hydrodynamics and small-scale shock physics: $E_\mathrm{iso} = 10^{53}$~erg, $n_{0} = 1$~cm$^{-3}$, $\epse = 0.35$, and $\epsB = 10^{-3}$. Comparing the NT-only and NL models in Figure~\ref{fig:CF2012_data} suggests that attempts to fit an afterglow model to these observations will arrive at very different values for these parameters depending on whether or not thermal electrons are present; this is seen directly in \citet{ResslerLaskar2017}, where determined parameters varied by a factor $1/f$ (where the participation fraction $f$ may be as small as 0.01) from their true values.  As such, what constitutes ``typical'' for a GRB may well change when hot thermal electrons are self-consistently included in models.

This work, \citet{WEBN2017}, and \citet{ResslerLaskar2017} present strong evidence that thermal particles present in the downstream plasma of GRB afterglows affect photon production and absorption at all energies, from GHz to GeV and above.  Future observations of afterglows, and attempts to infer physical parameters of the bursts, must consider this population.

\acknowledgments 
D.W. thanks Kate Alexander for helpful comments about the current state of GRB afterglow observations.  S.N. wishes to acknowledge the support of the Mitsubishi Foundation, the Associate Chief Scientist Program of RIKEN, and the Programs of Interdisciplinary Theoretical Science (iTHES) and Interdisciplinary Theoretical \& Mathematical Science (iTHEMS) at RIKEN.  B.M.V. acknowledges NSF grant AST-1306672, DoE grant DE-SC0016369 and NASA grant 80NSSC17K0757.

% bbbb  Note: must have files:  aa.bst  and  aa.cls
\bibliographystyle{aa} % A&A style

\appendix

\section{Analytical model for NT-only photon emission}
\label{app:analytical}

In Figure~\ref{fig:lowE_spectra_anCRTPNL} we compared four models: one analytical and three numerical.  In this appendix we motivate and explain the analytical model.  As in the NT-only model, we assume that not all electrons participate in photon production or absorption, and that those electrons that do participate form a power-law distribution.  To handle this we introduce a participation fraction $f$ \citep[as did][but the following discussion will show that our assumptions lead to markedly different results]{EichlerWaxman2005}.

Our electron distribution has a firm lower limit in energy, set by the large-scale hydrodynamics and robust microphysics:
\begin{equation}
  E_\mathrm{min} = \epse \gamZ m_{p} c^{2} .
  \label{eq:here_Emin}
\end{equation}
This is higher by a factor of a few\footnote{Specifically, $(p+1)/(p-2)$ for an electron spectral index $p$.} than the value for $E_\mathrm{min}$ typically used.  Since the spectral index of the power-law distribution is fixed to the test-particle value \citep{KeshetWaxman2005}, the number density of electrons in the distribution does not match the density required for charge neutrality.  These excess electrons would ordinarily fall in the thermal peak of the mixed thermal/nonthermal distribution.  As we are consciously excluding the thermal peak in our analytical model, we are faced with two choices, neither of which is physically correct.  We may either keep the minimum energy and ignore the (vast) majority of electrons needed for charge balancing, or keep the correct electron density and allow our distribution to extend to unrealistically low energies.  We choose the former option here, allowing only a fraction $f < 1$ of electrons to take part in photon processes.

To adhere to the Blandford-McKee solution, the remaining electrons (which, recall, are still at $\sim$GeV energies) are still assumed to be present in the downstream medium.  They do not, however, take part in any photon process.  This is not equal to assuming that they somehow avoid energization by plasma instabilities and enter the downstream plasma with very little energy.  Such cold electrons would nonetheless contribute to synchrotron self-absorption, as shown in \citet{ResslerLaskar2017}.  Our decision to ignore these electrons is, of course, unphysical, but it allows us to compare the following analytical model to the NT-only model and discuss the differences introduced by a more exact treatment of photon production and absorption.

The analytical model requires seven inputs: the peak flux density $F_{\nu,\mathrm{max}}$; the three break frequencies $\nu_{a}$, $\nu_{m}$, and $\nu_{c}$; and three spectral indices corresponding to the slopes of the spectrum in the segments below $\nu_{c}$.  Neither $\nu_{c}$ nor the three spectral indices are affected by our assumption of reduced electron participation, and so they retain the values presented in the literature \citep{SPN1998,GPS1999ApJ513,GPS1999ApJ527,Gao_etal_2013}:
\begin{equation}
  F_{\nu} \propto
    \begin{cases} 
      \nu^{2}  &  \nu < \nu_{a} \\
      \nu^{1/3}  &  \nu_{a} < \nu < \nu_{m} \\
      \nu^{-(p+1)/2}  &  \nu_{m} < \nu < \nu_{c}
    \end{cases}
  \label{eq:app_spec_indices}
\end{equation}
and
\begin{equation}
  \nu_{c} = 2.6\xx{16}~\mathrm{Hz}~(1+z)^{1/2} E_{52}^{-1/2} n_{0,0}^{-1} \epsilon_{B,-2}^{-3/2} t_{5}^{-1/2} .
  \label{eq:app_nuc}
\end{equation}
In the above definition for $\nu_{c}$, $z$ is the redshift, $E$ is the (isotropic equivalent) explosion energy of the GRB in ergs, $n_{0}$ is the upstream electron density in cm$^{-3}$, $\epsB$ is the energy fraction of the downstream plasma in the form of magnetic fields, and $t$ is the (observer frame) time in seconds.  Quantities with an additional numerical subscript are scaled as $Q = Q_{x} \cdot 10^{x}$.

The remaining quantities ($F_{\nu,\mathrm{max}}$, $\nu_{m}$, and $\nu_{a}$) are all affected by our assumptions and differ from their traditional forms.  The simplest change is to $F_{\nu,\mathrm{max}}$: fewer electrons participating in the production of photons means fewer photons produced, and so
\begin{equation}
  F_{\nu,\mathrm{max}} = 11~\mathrm{mJy}~(1+z) f E_{52} n_{0,0}^{1/2} \epsilon_{B,-2}^{1/2} D_{28}^{-2} ,
  \label{eq:app_Fmax}
\end{equation}
where $f$ is the participation fraction and $D$ the comoving distance to the GRB.  Since the minimum electron energy is defined by Equation~\ref{eq:here_Emin}, the value of $\nu_{m}$ no longer matches that found in the literature.  Instead, it becomes
\begin{equation}
  \nu_{m} = 4.93\xx{12}~\mathrm{Hz}~(1+z)^{1/2} E_{52}^{1/2} \epsilon_{e,-1}^{2} \epsilon_{B,-2}^{1/2} t_{5}^{-3/2},
  \label{eq:app_num}
\end{equation}
where $\epse$ is the fractional energy density carried by electrons at the shock front.  Note that, since $\epse$ enters the formula by way of the minimum electron energy $E_\mathrm{min}$, it implicitly counts the thermal (but non-radiating) population of electrons.  Finally, with fewer electrons participating in the synchrotron self-absorption process, there is less absorption.  The self-absorption frequency is thus
\begin{equation}
  \nu_{a} = 9.95\xx{6}~\mathrm{Hz}~(1+z)^{-1} f^{3/5} \epsilon_{B,-2}^{1/5} n_{0,0}^{3/5} \epsilon_{e,-1}^{-1} E_{52}^{1/5}.
  \label{eq:app_nua}
\end{equation}

\end{document}